\documentclass[submission,copyright,creativecommons]{eptcs}
\providecommand{\event}{TTC 2011} 
\usepackage{breakurl}             
\usepackage{graphicx}

\title{Saying Hello World with {MOLA} - A Solution to the {TTC} 2011 Instructive Case}

\author{Elina Kalnina 
\and 
Audris Kalnins
\and
Agris Sostaks
\and 
Janis Iraids
\and  
Edgars Celms 
\institute{Institute of Mathematics and Computer Science, University of Latvia,
\\ Raina bulvaris 29, LV-1459, Riga, Latvia} 
\email{\{elina.kalnina, audris.kalnins, agris.sostaks,
janis.iraids, edgars.celms\} @lumii.lv} }

\def\titlerunning{Hello World with MOLA}
\def\authorrunning{E. Kalnina, A. Kalnins, A. Sostaks, J. Iraids, E. Celms}
\begin{document}
\maketitle

\begin{abstract}
This paper describes the solution of Hello World transformations in MOLA transformation language. Transformations implementing the task are relatively straightforward and easily inferable from the task specification. The required additional steps related to model import and export are also described.\end{abstract}

\section{Introduction}

In this paper we describe the solution to Hello World case \cite{helloworldcase} for TTC 2011\footnote{\url{http://planet-research20.org/ttc2011/}} contest, implemented in MOLA model transformation language. The core task and all optional extensions are implemented. 
The {SHARE} image of solution is also provided. \cite{shareMOLA_TTC2011}

The ''Hello World'' task can be implemented in MOLA in a very straightforward way. We describe in the paper the basic principles of the solution. Before that, the situation with metamodels is described in some details, namely, how the metamodels are imported and extended and what implications this creates for source model import and target model export. The whole solution process using MOLA is briefly described as well. 

\section{MOLA Environment}
MOLA \cite{bib_mola} is a graphical transformation language developed at the University of Latvia. It is based on traditional concepts among transformation languages: pattern matching and rules defining how the matched pattern elements should be transformed. The formal description of MOLA and also the MOLA tool can be downloaded in \cite{webmola}.

A MOLA program transforms an instance of a source metamodel (defined in the MOLA metamodelling language - MOLA MOF, close to EMOF) into an instance of a target metamodel.

\emph{Rule} contains a declarative pattern that specifies instances of which classes must be selected and how they must be linked. Pattern in a rule is matched only once. The action part of a rule specifies which matched instances must be changed and what new instances must be created. The instances to be included in the search or to be created are specified using \emph{class elements} in the MOLA rule. The traditional UML instance notation (instance\_name:class\_name) is used to identify a particular class element. Class elements may contain constraints and attribute assignments defined using simple OCL-like expressions. Additionally, the rule contains association links between class elements. Class elements matched in one rule may be referenced in another one using the reference element (prefixed with the ''@'' symbol).

In order to iterate through a set of the instances MOLA provides the \emph{foreach loop} statement. The loophead is a special kind of a rule that is used to specify the set of instances to be iterated over. The pattern of the \emph{loophead} is given using the same pattern mechanism used by an ordinary rule, but with an additional important construct. It is the \emph{loop variable} - the class element that determines the execution of the loop. The foreach loop is executed for each distinct instance that corresponds to the loop variable and satisfies the constraints of the pattern. In fact, the loop variable plays the same role as an iterator in classical programming languages. The execution order in MOLA is specified in a way similar to UML activity diagrams.
MOLA has an Eclipse-based graphical development environment (MOLA tool \cite{webmola}, incorporating all the required development support. A transformation in MOLA is compiled via the low-level transformation language L3 \cite{bib_lx} into an executable Java code which can be run against a runtime repository containing the source model. 
For this case study Eclipse EMF is used as such a runtime repository, but some other repositories can be used as well (e.g. JGraLab \cite{kahl06}). 

\section{General Principles of Hello World Solution with MOLA}

The transformation development in MOLA starts with the development of metamodels. The MOLA tool has a facility for importing existing metamodels, in particular, in EMF (Ecore) format. Though MOLA metamodelling language (MOLA MOF) is very close to EMOF, and consequently Ecore, there are some issues to be solved. The current version of MOLA requires all metamodel associations to be navigable both ways (this permits to perform an efficient pattern matching using simple matching algorithms). Since a typical Ecore metamodel has many associations navigable one way, the import facility has to extend the metamodel. Another issue is the variable coding of references to primitive data types.

Metamodel import facilities in MOLA are able to perform all these adjustments automatically. This way the provided metamodels were imported into MOLA tool.

In some tasks for transformation development in MOLA additional metamodel elements are required, for example, in migration tasks to store relations between the source and target models. These metamodel elements have to be added manually. In the migration tasks, these are the associations between node classes in different graph encodings. Then the transformation itself (MOLA procedures) can be developed. The key features of transformations are described in the next section. The development ends with MOLA compilation.

Since the metamodels have been modified during import, the original source model does not conform directly to the metamodel in the repository, mainly due to added association navigability. Therefore a source model import facility is required. MOLA execution environment (MOLA runner) includes a generic model import facility, which automatically adjusts the imported model to the modified metamodel. Now the transformation can be run on the model. Similarly, a generic export facility automatically strips all elements of the transformed model which does not correspond to the original target metamodel.  Thus a transformation result is obtained which directly conforms to the target metamodel.
(For an inplace transformation the source and target metamodels coincide, as a result nothing has to be stripped.)
The transformation user is not aware of these generic import and export facilities, he directly sees the selected source model transformed. 

\section{MOLA Transformations for Hello World}
\begin{figure*}
	\centering
	\includegraphics[scale=0.8]{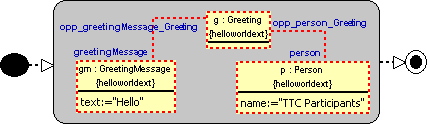}
	\caption{Solution of extended greeting creation task.}
\label{fig:p12}
\end{figure*}

The Hello world case consists of several very simple tasks. The first group of tasks are ''Greeting'' transformations. In these transformations the MOLA pattern used is very similar to the corresponding instance diagram given in the task specification. Solution of the second subtask is presented in Figure~\ref{fig:p12}. The grey rounded rectangle represents a MOLA rule. Class elements with red dashed border represent instance creation (similarly for association links).

\begin{figure*}
	\centering
	\fbox{\includegraphics[scale=0.8]{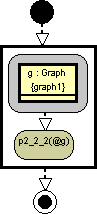}}
	\includegraphics[scale=0.8]{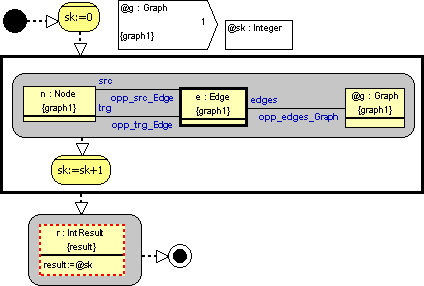}
	\caption{Two MOLA procedures: iteration through graphs on the left and transformation counting looping edges in a graph on the right.}
\label{fig:p22}
\end{figure*}

Another group of tasks is various kinds of instance counting in a graph. To solve these tasks an integer variable ''sk'' (the counter - a white rectangle in Figure~\ref{fig:p22}) and a MOLA foreach loop (a rectangle with bold border in Figure~\ref{fig:p22}) is used. In most cases the loophead pattern directly specifies the set of instances to be counted. Each loop iteration increases the instance count by one. After the loop the result instance is created and the counter value is placed into it. An example of such type is given on the right side of Figure~\ref{fig:p22}. In this case looping edges are counted. As it has not been defined in the task specification whether there is only one graph in a model, we assume that there can be many, therefore we process each graph separately. The counting procedure receives the graph to be processed as a parameter. The graph processing is given on the left side of the Figure~\ref{fig:p22}. This procedure calls the transformation counting looping edges for each graph (the procedure named ''p2\_2\_2''). A similar graph processing is done for all tasks where the phrase ''in a graph'' is used.

The only counting task processed differently is the circle counting. In MOLA there are two loop types: the foreach loop and while loop (rule + appropriate control flow). In the while loop, to ensure only distinct matches, an explicit marking of the already found matches (using a NAC construct) is required. This requires usage of temporary metamodel elements to solve the task. An alternative is to use three nested foreach loops since multiple loop variables are not supported in MOLA.  For details of both solutions see the appendix. To improve usability of MOLA in similar cases we could introduce a foreach loop with multiple loop variables. In this case the execution semantics would be similar to the one with nested loops.

\begin{figure*}[htbp]
	\centering
	\includegraphics[scale=0.8]{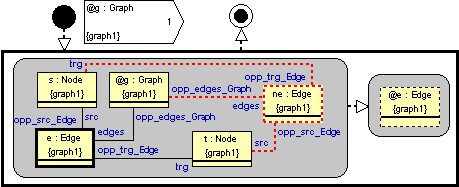}
	\caption{Transformation reversing edges in a graph.}
\label{fig:p3a}
\end{figure*}

\begin{figure*}[h!]
	\centering
	\includegraphics[scale=0.8]{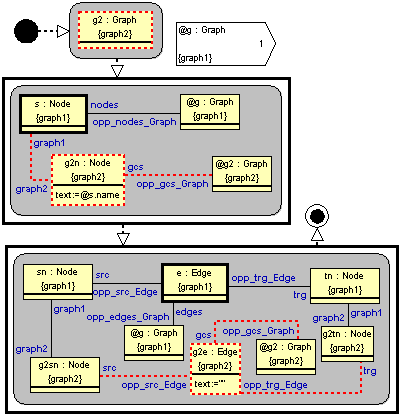}
	\caption{Model migration transformation. Migrates graph from encoding \emph{graph1} to encoding \emph{graph2}.}
\label{fig:p41}
\end{figure*}

The next task is to reverse edges in a graph. We propose to solve this task by adding a new reversed edge and removing the existing edge. This transformation is shown in Figure~\ref{fig:p3a}.

For model migration tasks additional temporary metamodel elements typically are required in MOLA. In this case it is sufficient to have an association between nodes in different graph encodings. After that the transformation is quite straightforward. At first we migrate nodes and then edges. Figure~\ref{fig:p41} shows the transformation.

The last group of tasks is deletion tasks. They are straightforward - the element to be deleted is found and then deleted.

We have also solved all optional tasks. However during the workshop it came up that we have misunderstood the task ''insertion of transitive edges''. The complete set of transformation procedures with short description is given in the appendix.

\section{Conclusions}
This case study has been quite appropriate to be implemented in MOLA. It confirms the assertion that simple tasks can be solved in a straightforward and easy readable way in MOLA. In most cases the basic part of the task is performed by one rule (or loophead). The used patterns have a natural form similar to what would appear in typical graph transformation languages. However, a natural solution of the circle counting task would require an even more powerful foreach loop feature in MOLA - a loophead with three loop variables. 

{\bf{Acknoledgments.}}
This work has been partially supported by the European Social Fund within the project "Support for Doctoral Studies at University of Latvia".

\bibliographystyle{eptcs}
\bibliography{HelloWorld}

\pagebreak
\section*{Appendix A}

In this appendix the complete set of transformation sources will be described. 

\begin{figure*}[htbp]
	\centering
	\includegraphics[scale=0.8]{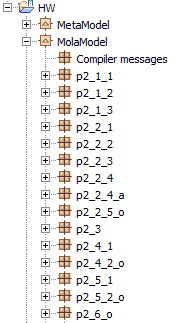}
	\caption{Structure of Hello World transformation project.}
\label{fig:aKoks}
\end{figure*}

The Hello World task consists of several small tasks. Therefore we have created one project with a package (compilation unit) for each sub-task. Content of the transformation project can be seen in Figure~\ref{fig:aKoks}. Each package contains a solution of one sub-task. ''pX\_Y'' means that the sub-task was defined in section X.Y of the case description. In some sections several sub tasks have been defined. In this case the third digit is used to describe the sub-task order in section. Optional sub-tasks are marked using ''\_o''. For one task two solutions are provided, the alternative solution is marked using ''\_a''.

\begin{figure*}[htbp]
	\centering
	\includegraphics[scale=0.8]{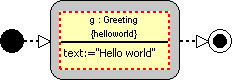}
	\caption{Transformation creating constant Greeting instance.}
\label{fig:ap11}
\end{figure*}

\begin{figure*}[htbp]
	\centering
	\includegraphics[scale=0.8]{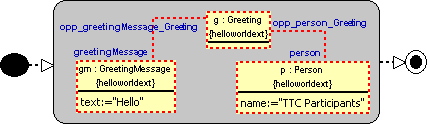}
	\caption{Transformation creating constant Greeting instance with references.}
\label{fig:ap12}
\end{figure*}

\begin{figure*}[htbp]
	\centering
	\includegraphics[scale=0.8]{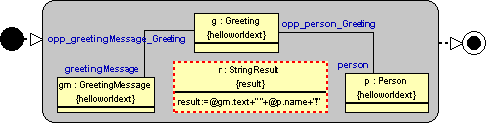}
	\caption{Model to text transformation creating greeting message.}
\label{fig:ap13}
\end{figure*}

Greeting transformations are given in Figure~\ref{fig:ap11}, Figure~\ref{fig:ap12} and Figure~\ref{fig:ap13}.
For these taks the transformation logic is described using one MOLA rule (grey rounded rectangle). In the first two tasks it is only required to create elements (marked with red dashed lines). In the third task an instance of class ''StringResult'' is created, if the pattern (elements with black solid lines) in MOLA rule is matched.

\begin{figure*}[htbp]
	\centering
	\fbox{\includegraphics[scale=0.8]{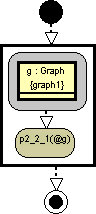}}
	\includegraphics[scale=0.8]{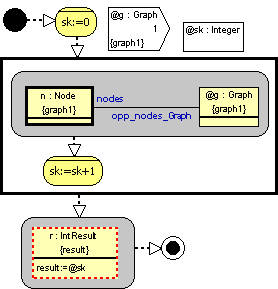}
	\caption{Transformation counting nodes in a graph.}
\label{fig:ap21}
\end{figure*}

\begin{figure*}[htbp]
	\centering
	\includegraphics[scale=0.8]{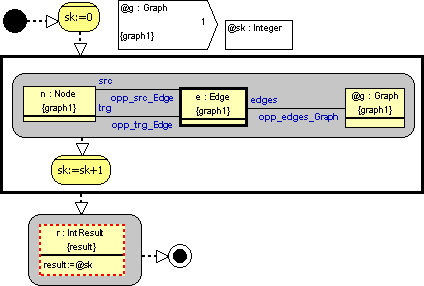}
	\caption{Transformation counting looping edges in a graph.}
\label{fig:ap22}
\end{figure*}

\begin{figure*}[htbp]
	\centering
	\includegraphics[scale=0.8]{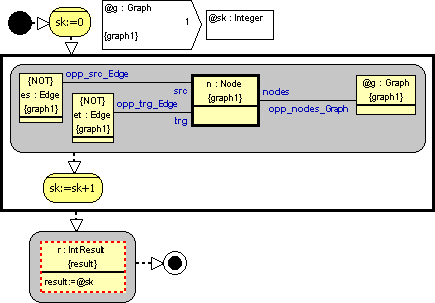}
	\caption{Transformation counting isolated nodes in a graph.}
\label{fig:ap23}
\end{figure*}

The next group of tasks in the task specification is instance counting tasks. A transformation counting nodes in a graph is given in Figure~\ref{fig:ap21}. A transformation counting looping edges is given in a Figure~\ref{fig:ap22}. The transformation counting isolated nodes is given in Figure~\ref{fig:ap23}. In MOLA the counting is implemented using an integer counter and a foreach loop where the counter is increased. 
A MOLA variable (white rectangle) of type integer is used as a counter. To modify the values of counter text statements (yellow rounded rectangles) are used. Finaly to save the counting result in the resulting model a MOLA rule creating an instance of class ''IntResult'' is used.

\begin{figure*}[htbp]
	\centering
	\includegraphics[scale=0.8]{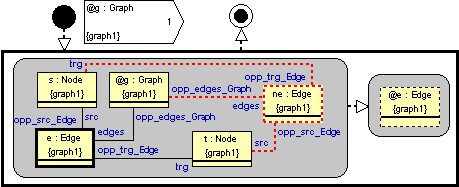}
	\caption{Transformation inversing edges.}
\label{fig:ap3a}
\end{figure*}

\begin{figure*}[htbp]
	\centering
	\includegraphics[scale=0.8]{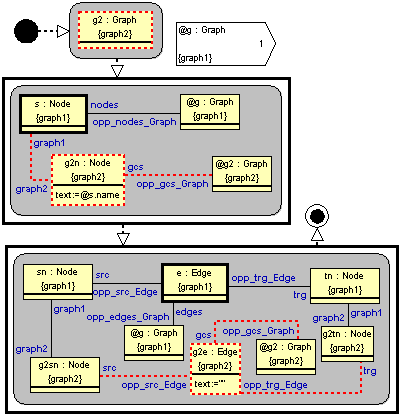}
	\caption{Model migration transformation. Migrates graph from encoding \emph{graph1} to encoding \emph{graph2}.}
\label{fig:ap41}
\end{figure*}

\begin{figure*}[htbp]
	\centering
	\includegraphics[scale=0.8]{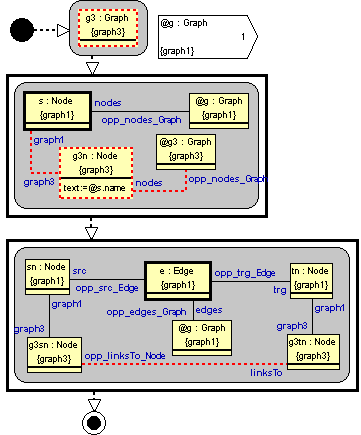}
	\caption{Solution of optional model migration task. Migrates graph from encoding \emph{graph1} to encoding \emph{graph3}.}
\label{fig:ap42o}
\end{figure*}

\begin{figure*}[htbp]
	\centering
	\includegraphics[scale=0.8]{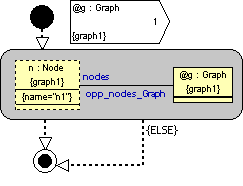}
	\caption{Transformation that deletes node named ''n1'' (if such node exists) in a graph.}
\label{fig:ap51}
\end{figure*}

\begin{figure*}[htbp]
	\centering
	\includegraphics[scale=0.8]{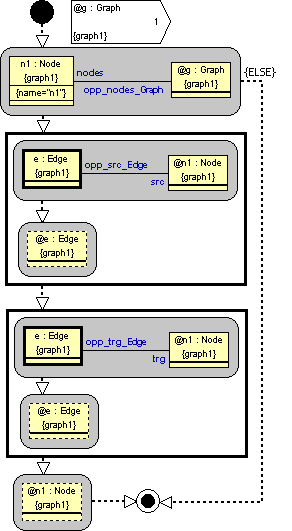}
	\caption{Transformation that deletes node named ''n1'' (if such node exists) and its incident edges in a graph.}
\label{fig:ap52o}
\end{figure*}

\begin{figure*}[htbp]
	\centering
	\includegraphics[scale=0.8]{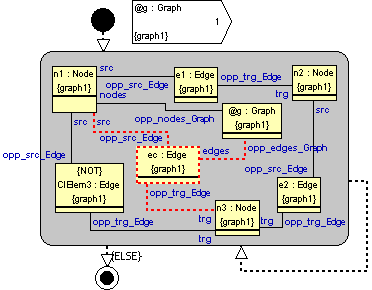}
	\caption{Solution of optional task: insertion of transitive edges.}
\label{fig:ap6o}
\end{figure*}

For all these tasks it was required to count elements in a graph. However it was not defined whether the model contains only one graph or multiple. For transformations to work properly when there is more than one graph in a model we provide the graph to be processed as a parameter. 
Therefore we use another MOLA procedure where we iterate through all graphs in a model (using a foreach loop) and from here we call the transformation (using call statment) procesing the current graph. 
An example of such transformation is given on the left side of Figure~\ref{fig:ap21}. (The only thing that changes is the called procedure.) If there is always only one graph in a model this step could be omitted. The same could be said also about transformations in Figure~\ref{fig:ap3a}~-~\ref{fig:ap6o}.

The next task was to count circles consisting of three nodes. The solution of this task is different from the previous one because we want to find all different circles. In this case one loop variable is not sufficient and as a result, several loops are required.

\begin{figure*}[htbp]
	\centering
	\includegraphics[scale=0.8]{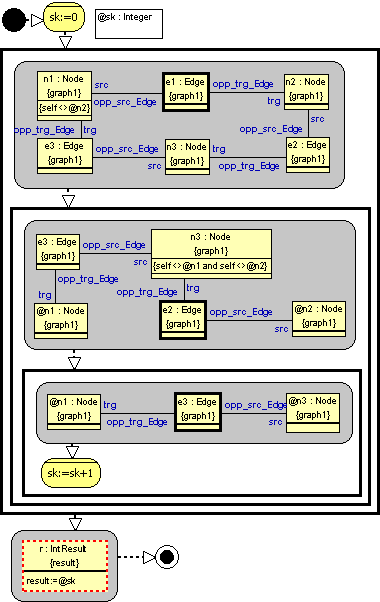}
	\caption{Transformation counting circles consisting of three nodes.}
\label{fig:ap24}
\end{figure*}

In the task specification it was not clearly stated whether graphs or multi-graphs should be considered (i.e., is it possible to have multiple edges between two nodes.) As the provided metamodel supports multi-graphs and graphs are a subclass of multi-graphs, we decided to build our solution so that multi-graphs are supported. As we support multi-graphs, if there is a circle ''n1;n2;n3'' and two edges between ''n1'' and ''n2'', then there will be two circles ''n1;n2;n3'' (and 2*''n2;n3;n1'' + 2*''n3;n1;n2''). Solution of this task is given in Figure~\ref{fig:ap24}. As we want to distinguish different edges between the same nodes, edges are used as loop variables. There are three nested loops used in the solution. Each loop selects one edge for the circle. Actually finding of circles is defined in the loophead of the first loop, however using this loop we are only able to find all edges which are part of some circle, but we don't know in how many circles this edge is used. Adding the second and third loop we count all circles consisting of different edges 3 times as required in the task specification.

If we know that there are no multi-graphs then the last loop can be omitted, because the existence of the third edge is already validated by patterns in the first and second loop. However in this case it is probably easier to understand if nodes are used as loop variables, but for that again three loops are needed.

\begin{figure*}[htbp]
	\centering
	\includegraphics[scale=0.8]{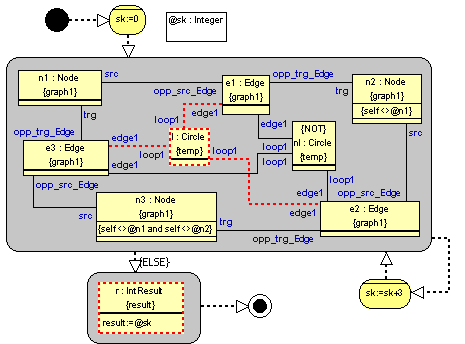}
	\includegraphics[scale=0.8]{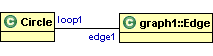}
	\caption{Transformation counting circles consisting of three nodes using temporary metamodel elements.}
\label{fig:ap24a}
\end{figure*}

Solution of this task is quite lengthy, however if we add temporary classes it is possible to create a shorter and more elegant solution. We extend the metamodel by adding a temporary class ''Circle'' and connecting it to the class ''Edge''. The metamodel extension is shown on bottom of Figure~\ref{fig:ap24a}. If such extended metamodel is used then we can simply write a MOLA rule looking for circles and marking the found circles: connecting all edges of a circle to a new instance of the ''Circle'' class. To ensure that each circle is found exactly once a NOT constraint (an equivalent to NAC in graph transformation languages) is used stating that this circle hasn't been marked previously. As in this solution we don't care about the order of edge finding, the loop counter is increased by 3, to ensure that each circle has been counted three times. This solution is given in Figure~\ref{fig:ap24a}.

\begin{figure*}[htbp]
	\centering
	\includegraphics[scale=0.8]{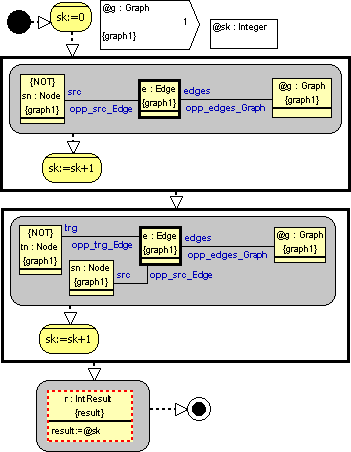}
	\caption{Solution of optional task: counting of dangling edges.}
\label{fig:ap25o}
\end{figure*}

Next was an optional task to count dangling edges. The solution is given in Figure~\ref{fig:ap25o}. In this case two loops are used. The first one counts edges without a source. To ensure that edges without source and without target are counted only once the second loop counts only edges with a source and without target.

The next task we consider is edge reversing. We selected a solution where a new reverted edge is created and the old edge is deleted 
(delete is marked using a black dashed line). 
The solution is displayed in Figure~\ref{fig:ap3a}. Actually a shorter solution in MOLA is possible; however it is not supported by the current version of MOLA tool.

The next group of tasks was model migration tasks. To implement such tasks it is necessary to add temporary traceability relations to the metamodel. In this case it is sufficient to have an association between nodes in both metamodels. The migration transformation from the metamodel \emph{graph1} to metamodel \emph{graph2} is given in Figure~\ref{fig:ap41} and from metamodel \emph{graph1} to metamodel \emph{graph3} in Figure~\ref{fig:ap42o}. In both cases at first a new graph in the target model is created. After that all nodes are cloned and traceability links added. 
(For this a foreach loop iterating through all nodes in the source graph is used.) 
Finally all edges are transformed using the traceability information to find the appropriate source and target nodes in the migrated model.
(For this a foreach loop iterating through all edges in the source graph is used.) 

The last mandatory transformation is deletion of the node named ''n1''. This transformation is very straightforward (see Figure~\ref{fig:ap51}). We try to find such node using a MOLA pattern and if it is found we delete it. The deletion is represented by a black dashed line. In the extension it was required to delete all incident edges as well. The solution of extension is given in Figure~\ref{fig:ap52o}. In this case at first the node is found, then all outgoing edges are deleted, after that all incoming edges are deleted and finally the node itself is deleted.

The last task in the case description is insertion of transitive edges. The easiest way to solve this task in MOLA is using a while loop. While it is possible to insert a transitive edge the MOLA rule is executed repeatedly, after that the end is reached (see Figure~\ref{fig:ap6o}). However, this solution computes the transitive closure instead of $R^2$ required in the task specification. To compute $R^2$ the newly created edges should be marked and a filter to exclude them from matching in the rule should be added.

\end{document}